\documentclass[twocolumn,preprintnumbers,amsmath,amssymb,superscriptaddress]{revtex4-1}
\usepackage{graphicx}
\usepackage{color}
\usepackage{amsmath}
\usepackage{latexsym}
\usepackage{epstopdf}
\usepackage{amsmath}
\usepackage{amssymb,amsmath}
\usepackage{dcolumn}% Align table columns on decimal point
\usepackage{bm}
\setcitestyle{super}
\begin{document}
\title{Fast water flow through graphene nanocapillaries: a continuum model approach involving the microscopic structure of confined water }
%%\title{Melting of mono-layer fluorographene: a molecular dynamics study}
%\author{}
\author{M. Neek-Amal$^{1,2,6}$,  A. Lohrasebi$^3$, M. Mousaei$^1$, F. Shayeganfar$^4$,  B. Radha$^{5,6}$,  and F. M. Peeters$^{2,6}$ \\
\small$^1$Department of Physics, Shahid Rajaee Teacher Training  University, 16875-163, Lavizan, Tehran, Iran.\\
\small$^2$Department of Physics, Universiteit Antwerpen, Groenenborgerlaan 171, B-2020 Antwerpen, Belgium.\\
\small $^3$Department of Physics, University of Isafahan, 81746-73441, Isfahan, Iran.\\
\small $^4$Department of Civil and Environmental Engineering, Rice University, Houston, Texas 77005, USA\\
\small$^5$School of Physics and Astronomy, University of Manchester, Manchester M13 9PL, United Kingdom.\\
\small$^6$National Graphene Institute, University of Manchester, Manchester, M13 9PL, United Kingdom.
 }
\date{\today}

\begin{abstract}
Water inside a nanocapillary becomes ordered, resulting in unconventional  behavior.  A profound enhancement of water flow inside nanometer thin  capillaries made of \texttt{\textbf{\emph{graphene}}} has been observed [B. Radha et.al., Nature (London) {\bf538}, 222 (2016)]. Here  we explain this enhancement as due to the large  density and the extraordinary viscosity of water inside the graphene nanocapillaries. Using the Hagen-Poiseuille theory with slippage-boundary condition {and  incorporating disjoining pressure term in combination with} results from molecular dynamics (MD) simulations, we present an analytical theory that elucidates the origin of the enhancement of water flow inside {hydrophobic}  nanocapillaries.
  Our work  reveals a distinctive dependence of water flow in a nanocapillary on the structural properties of nanoconfined water in agreement with experiment, which opens a new avenue in nanofluidics.
\end{abstract}

\maketitle

%\section{Introduction}
 Water flow through nanoscale channels, and  the determination of the slip length, have been the subject of intensive studies\cite{hendi,PNAS17,rev,falk,maj,Holtsci,Holt,Qin,hummer,Thomas,Rasaiah,maj,Holtsci,Holt,Qin,Lac,Noy,ber,jos,thom,alur,sokhan,quir}. In a recent study Radha \emph{et al.}\cite{Radha}  fabricated atomically flat 2D-capillaries and was able to control the water flux through the channel size.  In Ref. [17], an unexpectedly fast flow (up to 1 m/s) of water through flat nanochannels was reported\cite{Radha}. In addition to the large slip length, this unexpected phenomena might be due to the high disjoining pressures\cite{Israelachvili} inside the nanochannel. The disjoining pressure is added to the well-known capillary pressure that causes oscillation in the meniscus pressure which for channels thinner than H=8\AA,~was found to be in the order of 1\,kbar\cite{IEE,PRE16}.

In the continuum limit, transport of water through a capillary is described by the Hagen-Poiseuille equation (HPE), however, for nanofluidics  several modifications (beyond the no slip-boundary conditions) should be made\cite{Radha,Israelachvili,IEE,PRE16,Ahn,kh,khh,khhh,Ebrahimi}.
There have been several  studies on the ordering of water inside a {hydrophobic nanocapillary}\cite{Algara,Qiu,ACS16,Sobrino,Sobrino2}. Particularly  monolayer/bilayer ice confined within a {hydrophobic nanochannel} has been studied using MD simulations\cite{Qiu,ACS16,Sobrino,Zangi,Walt} and using density functional theory calculations\cite{Corsetti1}. Such an ordering of water molecules can change significantly the density\cite{Qin,hamid,Pradeep} and its viscosity\cite{ACS16}  inside the channel. {Using molecular dynamics  simulations the pressure-driven water flow through carbon nanotubes (CNTs) with diameters ranging from 0.83 to 1.66 nm were studied by Thomas et. al.\cite{PRL2009}, where a transition from continuum to subcontinuum flow with decreasing CNT diameter was found. While the standard linear relationship in Darcy law is violated, they modified the Darcy (continuum) equation in order to explain their molecular dynamics simulations results.\cite{PRL2009}}

%{On the other hand, the measured viscosity of water confined between hydrophilic layers (e.g. mica surfaces) can be a factor of three larger than its bulk value~\cite{Nature2001}. In fact, the viscoelastic properties of nanoconfined water change significantly with the %chemical nature of the confining walls and the dynamic state of the confined liquid, e.g. for hydrophilic confinement a low viscosity enhancement was observed at high shear rates~\cite{PRL1,PRL2,Nature2001} while many orders of magnitude enhancement was observed %at small shear rates ~\cite{JACS}.}

Here we will demonstrate the  profound influence of the density and viscosity of water inside graphene {nanocapillaries} on the water flow rate. Our calculations are based on the  well-known continuum model formalism but taking into account the  ordering of the water molecules. We propose an analytical model to describe experimental results of Ref. [17], i.e. fast water flow in {graphene nanocapillaries}, employing aforementioned microscopic structure of confined water. It is entropically unfavorable for a
hydrophobic surface to bind water molecules via ionic or hydrogen bonds resulting in low friction of water inside graphene capillaries.  The water-solid wall slip length {(at molecular scale)} is much larger than the capillary size which  results in different
boundary conditions as compared to bulk water in macroscopic channels. { It is well-known that  the slip dynamics appears through three different length scales: i) individual molecular, ii) beyond-few molecules, i.e. actual slip at a liquid-solid boundary, and iii) apparent slip due to the motion over complex boundaries.  %In the past decades some form of fast flow or equivalently reduced resistance to fluid motion, has been reported~\cite{MITreview}.}}
Using a slip length of about 600\AA,~and a contact angle close to 90$^o$, our analytical results  agree very well with  recent experiments on water flow through graphene nanochannels\cite{Radha}.

\begin{figure}[t]
\includegraphics[width=0.8\linewidth]{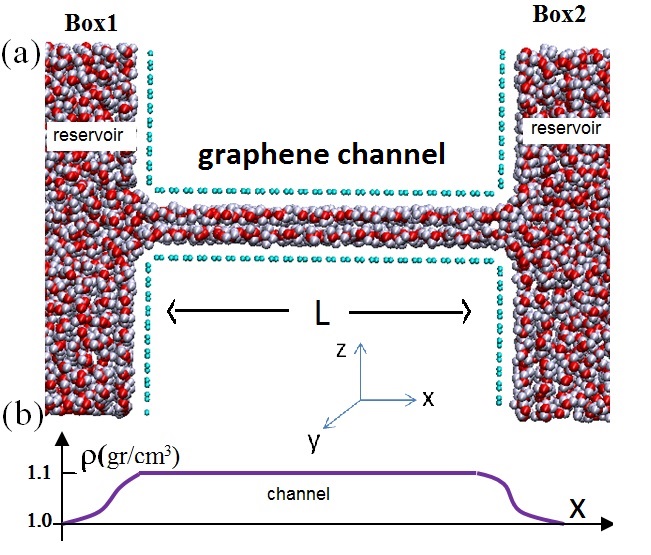}
\caption{ (color online) Side  view of a snapshot of water molecules between two graphene sheets separated by H=10\,\AA~ taken from our MD setup.~Red (gray) balls are oxygen (hydrogen).  (b) A schematic view of the variation of water density inside the reservoirs and the channel.
~\label{fig1}}
\end{figure}

\emph{The model.}
The Poiseuille flow solution using no-slip boundary condition for a channel with height $H$, which is subjected to a pressure difference ($\Delta$P) along the length of the channel, is quadratic in velocity. For water flow through $n$ equal  nanochannels  when the effect due to the slip velocity  ($\frac{\lambda du_\textrm{x}}{d\textrm{z}}$)  is taken into account, the volumetric flow rate is given by\cite{Holt}
\begin{equation}\label{Q}
Q=\rho \frac{|\Delta P|}{12\eta}  H^3 (1+\frac{6\lambda}{H}) \frac{ nw}{L},
\end{equation}
where $\lambda$ is the slip length, “$\rho$ is the density, $\eta$ is the viscosity, and $L$ is the length of each channel having a width $w$.
Notice that in Eq.~(\ref{Q}) the slip term is dominant  for $\lambda\gg H$. For water inside a nanochannel  the density $\rho(H)$ and viscosity $\eta(H)$  vary with the capillary size\cite{size1,size2,size3}.  The density  increases due to the fact that the accessible volume for water molecules in a nanochannel is smaller than the geometrical volume -- because of excluded volume effect near the confining walls. Furthermore,  the interaction within the layers and with the {hydrophobic confining walls} induces structuring in water which significantly enhances the viscosity\cite{ACS16,slabgeom,Pradeep}. Such changes in the density and viscosity are expected to be pronounced for  nanochannel with height H$\preceq$12\,\AA.~ We take $\rho(H)=\rho_0 f(z)$ and $\eta(H)=\eta_0g(z')$  where $\rho_0\cong1$\,gcm$^{-3}$ and $\eta_0\cong0.89$\,mPa.s are the bulk values for density and viscosity, respectively. Here $z=H/\delta$ and $z'=H/\delta'$
are exponential decay lengths for density and viscosity inside the channel {where the two parameters $\delta$ and $\delta '$ are determined from MD simulations.}
  The two functions should approach $f(z)\rightarrow 1,~g(z') \rightarrow 1$ when $H>>\delta,~\delta'$. We propose the following  functions fulfilling these boundary conditions
 \begin{equation}\label{dens}
f(z)=1+a e^{-z},~~~ ~g(z')=1+b e^{-z'},
\end{equation}
where a, b, $\delta$, and $\delta'$ will be obtained by fitting to results from our MD simulations.
\begin{figure}[t]
\includegraphics[width=.475\linewidth]{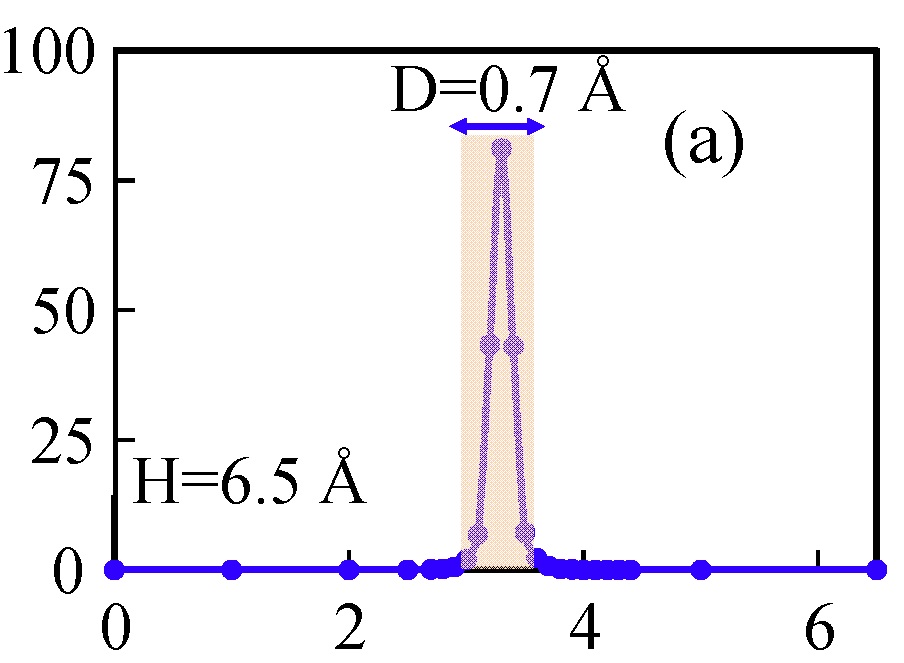}
\includegraphics[width=.475\linewidth]{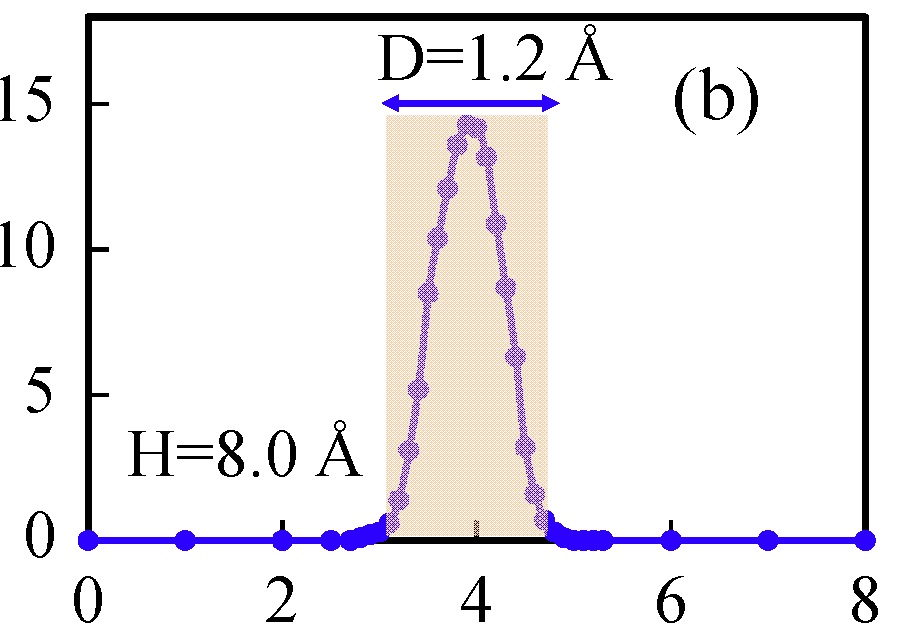}
\includegraphics[width=.475\linewidth]{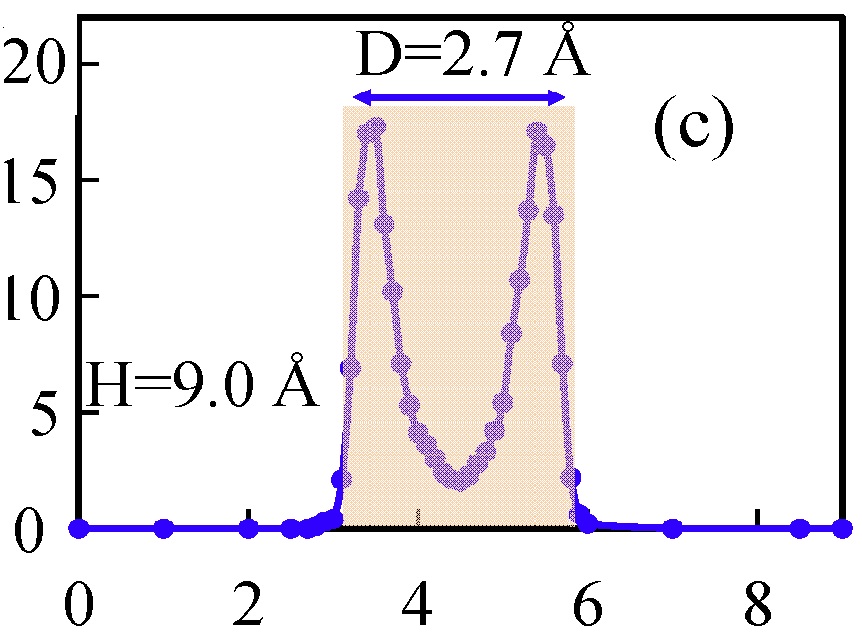}
\includegraphics[width=.475\linewidth]{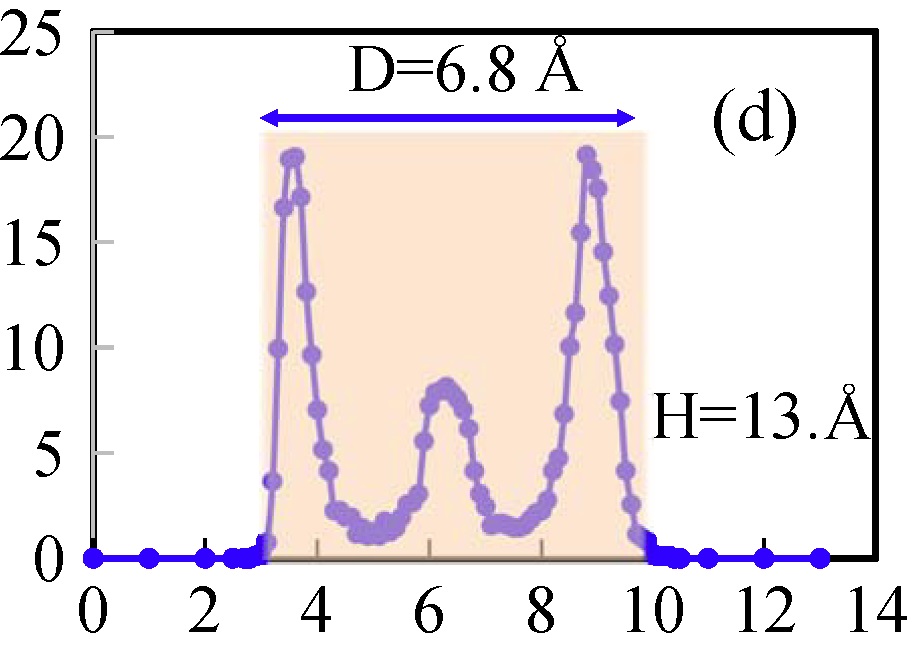}
\caption{ (color online)  Density profile perpendicular to the channel for four typical channel heights. The colored regions show the  accessible volume  for  water  which defines  the effective height D.
~\label{fig2}}
\end{figure}

\begin{figure}[t]
\includegraphics[width=0.80\linewidth]{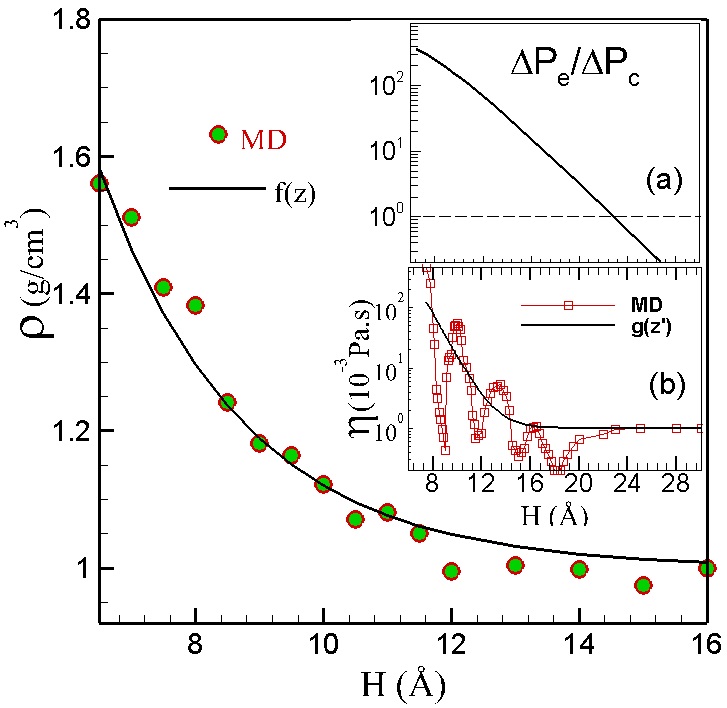}
\caption{ (color online) The density of water inside the nanochannel as function of the channel height H. The symbols are MD simulation results using the definition for the accessible volume for water molecules (see Fig. 2). The solid curve is a fit with Eq. (2) ($f(z)$).   In the inset (a), we show the variation of the ratio between entropic and capillary pressures ($\frac{\Delta P_e}{\Delta P_c}$) with channel size. The inset (b) shows the MD simulation results for viscosity (symbols) and corresponding $g(z')$ function (solid curve) according to Eq. (2).
~\label{fig3}}
\end{figure}

\begin{figure}[t]
\includegraphics[width=1.0\linewidth]{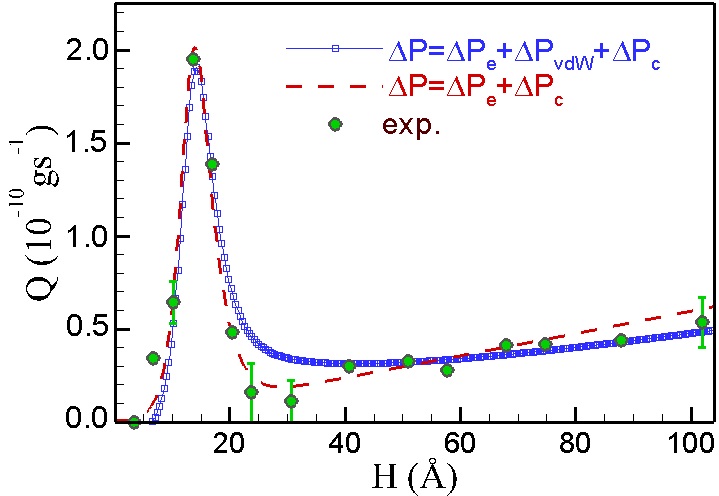}
\caption{ (color online) Water flow rate - Q - as function of channel height. The experimental data and corresponding error bars are shown by green-dots\cite{Radha}. The red-dash curve  is the theoretical result including capillary and entropic pressure when $\lambda$=600\,\AA.~ The blue-dotted curve is  water flow rate when also the vdW pressure is included, see Eq.~(4).
%The arrows refer to the data from single-device measurements. The green error bars indicate experimental detection limit.
~\label{fig4}}
\end{figure}

%\begin{figure}[t]
%\includegraphics[width=0.95\linewidth]{Fig5.png}
%\caption{ (color online) The variation of viscosity with channel size H. The solid and dash curves are viscosity of water inside nanochannel when $\delta=\delta'$ and $\delta\neq\delta'$, respectively where  $\lambda$=600\AA.~ The squares symbols are our molecular dynamics simulation results\cite{ACS16}.
%~\label{fig5}}
%\end{figure}

By assuming that the hydrostatic pressure is many orders of magnitude smaller than:  i) the Laplace pressure\cite{PRE16,Ebrahimi} which corresponds to the traditional capillary pressure $\Delta P_c=\frac{\gamma}{H}=\frac{2\sigma cos(\Phi)}{H}$ (here $\sigma\simeq70mN/m$ is the interfacial tension\cite{surfacetension} and $\Phi$ is the contact angle between water and graphite/graphene. The latter has been measured, however the results {vary widely which can be traced back to functional groups on the surface}, i.e. $\Phi \in[55^o-127^o]$~\cite{PNAS17,Taherian,Mucksch,zisman,nanolett2001}). ii) The disjoining pressure (DP) $\Delta P_d=-\frac{1}{A}(\frac{\partial G}{\partial H})_{T,V,A}$~\cite{DP2} which is due to the ordered structure of water inside the nanochannel\cite{Sobrino,ACS16,Algara} and the interaction of water with the channel wall. The DP can be one to three orders of magnitude larger than the capillary pressure and is a consequence of van der Waals (vdW) and entropic components\cite{IEE}.
Substituting aforementioned pressures in Eq.~(\ref{Q}), and by including both the density and viscosity functions, we find
\begin{equation}\label{QQ}
Q=A  \frac{f(z)}{g(z')} (H^2+6 H\lambda) [\gamma+H \Delta P_d].
\end{equation}
The parameter A=$\frac{nw\rho_0}{12\eta_0 L}$ is taken as a scaling factor in our model (using experimental numbers $w=$1300\AA~ and $L=10^4$\AA~for one channel, is about 1.2172$\times10^{-4}$ s\AA$^{-2}$ which is very close to our obtained number from a fitting on the experimental data, i.e. 1.2174$\times10^{-4}$ s\AA$^{-2}$).
By measuring the chemical potential difference inside the capillary and the reservoirs, the entropic pressure is given by\cite{Qin} $\Delta P_e=\frac{RT}{V_m} ln⁡(f(z))$ where RT=2494.2\,J/mol, and $V_m$=18\,cm$^3$/mol.

%Before finding $f(z)$ and $g(z')$ we discuss the possibility of large densities for confined water.
Computer simulations and experiments confirmed the existence of an ordered structure for confined water inside capillaries with H$\preceq12$\,\AA.\cite{Algara,Zangi,ACS16,Sobrino,Corsetti1}.
%~Confined water has a net zero electric dipole and the dipoles of the water molecules are in the same plane resulting in planar ice.\cite{Algara,Sobrino2} The distance between confined water molecules is  in the range of [2.7-2.9]\,\AA~which is the same as for bulk water. The latter information immediately helps us to calculate the density of nanoconfined water with  a semi-square lattice structure yielding $\rho_{2D}=\frac{m}{V}=[1.52-1.22]$\,gcm$^{-3}$ by using the aforementioned $d_{OO}$ distance ($m$=18\,g/mol and $V$=$d_{OO}^3$).
The distance $d_{OO}=$2.8\,\AA~ in semi-squared lattice structure of confined water results in a water density of about~1.36\,gcm$^{-3}$.

{We performed extensive MD } simulations using TIP3P force field\cite{tip3p} to find the density  and  the structure of water inside nanochannels.~Our simulation setup consists of three elements, i.e. a graphene nanochannel {elongated in the x-direction} having height H$\in$[6.5\,\AA~to~16\,\AA], {length $L=50$\,\AA,~ and width $w=20$\,\AA}~which connects two water reservoirs on both sides of the channel, see Fig.~\ref{fig1}.  The simulations are performed using LAMMPS package and we employed an NVT ensemble. The long-range electrostatic interactions were computed with the particle-particle particle-mesh (PPPM) method with a cutoff distance of 12\,\AA.~The non-bonding interactions were modeled by using the Lennard-Jones(LJ) potential. In Fig.~\ref{fig1}(a)  we show a side view of confined water and the two reservoirs for a channel with height H=10\,\AA.
%{Our simulation setup is more realistic than the previous setup\cite{Radha} where a specious water suction inside capillary were %reported. ---- We did not observe any suction of water when we repeated the simulations proposed by  F. C. Wang\cite{Radha}.}

First we calculate the density profile across the channels (i.e. perpendicular to the graphene layers, i.e. along the z-axis). {These profiles were calculated by counting the number of water molecules in the yz-plane (across the channel) and averaging along the channel, i.e. $\overline{N(\textrm{z})}=\int\int n(\textrm{x,y,z})d\textrm{y}d\textrm{x}/Lw$ where  $n(\textrm{x,y,z})$ is the number of molecules at point $(\textrm{x,y,z})$ inside the channel.} Four typical density profiles are shown in Fig.~\ref{fig2}. {We found that e.g. for H=6.5\,\AA,~ and 8\,\AA~ $\overline{N(\textrm{z})}$ are Gaussian functions with  decreasing width for decreasing H. For other H's, $\overline{N(\textrm{z})}$ is divided into two or more distinctive peaks. The middle peak (see Fig.~\ref{fig2}(d)) disappears by increasing H and only two peaks  remained close to the edges\cite{hamid}.~ Note that beyond H=11.0\,\AA~the number of layers and D  (where distance `$D$' is effective height of the channel with geometrical height H) becomes larger. For H=13\,\AA~the middle layer is smaller than those on both sides {which plays an important role in the fast water flow}.}~These results indicate that the accessible volume for water inside the nanochannels (grayed rectangles in Fig.~\ref{fig2}) is  smaller than the geometrical volume. Therefore, the accessible volume for water inside the channels is proportional to $\frac{N(H)}{D}$ instead of $\frac{N(H)}{H}$. Here $N(H)=\int \overline{N(\textrm{z})}d\textrm{z}$ is the total number of water molecules per surface area inside the channel with size H. In most of the cases, regions of about 3.0\,\AA~from both sides of the graphene walls are inaccessible for the  water molecules\cite{hamid} (independent of H).  They form an excluded volume due to the graphene wall-water interaction. Obviously, for larger H, the accessible volume approaches the geometrical volume, i.e. the relative difference $(1-\frac{D}{H}) \rightarrow 0$.

{Henceforth, the problem of determining the density is reduced to determining `D' and the corresponding volume. D is taken such that 98$\%$ of water molecules are confined in the middle of the channels (i.e. within the colored rectangles shown in Fig. 2). We found H-D is almost the same for H$\succeq12$\,\AA~ and the corresponding density is close to bulk water.~For the smallest H i.e. H=6.5\,\AA,~we obtain almost planar and square-rhombic lattice structure at room temperature and lateral pressure of about 0.9GPa with $d_{OO}=2.8\pm0.05$\AA,~ from which we determine a maximum density around 1.4\,gcm$^{-3}$.~This can be found only if we use D$\approx1.0\pm$0.3\AA.~Alternatively one may use the vdW radius of  O and C atoms to define the effective height\cite{arxiv2017,Pradeep}.~These led us to conclude that the density of confined water is larger than the bulk density which is in agreement with previous reports\cite{Pradeep,hamid}. The circle symbols in Fig.~\ref{fig3} are the densities found from our MD simulations using the aforementioned D values. The corresponding $f(z)$~function using the best fit of the MD data is shown by the solid line in Fig. 3 with $a$=10.9 and $\delta$=2.2\,\AA.~The profile shown in Fig.~\ref{fig1}(b) schematically shows the variation of the density inside and outside the channel with H=10\,\AA.%\cite{note4}}

Our approach is general and can also be used to describe fast water flow in CNTs. Using an array of field effect transistors defined on individual CNTs, Qin \emph{et al.}\cite{Qin} measured the water flow rate through individual CNTs and found  a rate enhancement of $\sim$882 for CNTs with diameter of 8.1\,\AA.~{The water density in the CNT can be described by the same function $f(z)=1+25.e^{-z}$ } where $z=\frac{r}{2.5}$ and $r$ is the {diameter in units of Angstrom.}

The dependence of the  viscosity on H was  reported in our previous work\cite{ACS16}.  In general viscosity is direction dependent in confined systems where the major contribution of the viscosity is due to its xy-component, see Ref. [20]. The large viscosity (in $H<13\AA$)~is due to the layered/ordered structure of water
%, i.e. for H$<$10\,\AA~it was found that $\eta(H)\sim100\eta_0$.
The MD simulation results are shown in Fig.~\ref{fig3}(b).  {A typical fit $g(z')$ (which satisfies the above mentioned boundary requirements) on our MD data is shown by the solid curve in Fig.~\ref{fig3}(b) with parameters b=6.23$\times10^4$ and $\delta'=1.19$\,\AA.~
%It is worth to mention that the viscosity can be a decreasing function of contact angle\cite{PNAS17}.

{Using the obtained $f(z)$ and $g(z')$, we first use $\Delta P_d=\Delta P_e$ (neglecting the\, vdW pressure). The scaling parameter A is the only fitting parameter and our theoretical results for Q are shown in Fig.~\ref{fig4} (red dashed curve) where the slip length was set to be $\lambda=$600\,\AA~ and $\gamma=$ $0.1$\,mN/m which results in a contact angle of 89.96$^o$ by using $\sigma=70mN/m$.   The symbols in Fig.~\ref{fig4}  are the experimental data\cite{Radha} (the green error bars indicate the experimental detection limit) and dash-red curve  is the calculated water flow rate. Our results are  in very good agreement with the experimental results.~In the inset of Fig.~\ref{fig3}(a), we depict the ratio between $\Delta P_e$ and  the capillary pressure which is larger than 40 for H$<$13\,\AA.~The capillary pressure is larger than the entropic pressure only for H$>$22\,\AA.~}

Finally, it is worthwhile to investigate the effect of vdW pressure.\cite{IEE} To that end, we add the vdW pressure to the DP as
\begin{equation}\label{QQ2}
  \Delta P_d=\frac{A_H}{6\pi H^3}+\Delta P_e,
\end{equation}
  {where $A_H\simeq35zJ$\cite{Hamaker} is the Hamaker constant for intercalated graphite with water. Using the previous $f(z)$ and $g(z')$ and $\lambda=600$\,\AA~we obtain  the blue-dotted curve in Fig.~\ref{fig4}. It is seen that, including vdW pressure  significantly improves the results for large H (H$>$80\AA) while the results for H= 23 and 30\AA~are slightly overestimated. Therefore disjoining pressure causes a substantial enhancement of  the water flux  for channel heights around 13\AA.~
Notice that~ previous MD~ simulations~(see Fig. 3 in Ref.~ [20] found peak in~ the~ water-flow ~rate~ that~ are~ located~ in~ the~[17-23]\AA~ range which deviates from the experiment where the peak is around 13\AA.~%
%Therefore the MD simulation results in Ref. [7] do not  quantitatively  support experiment.

%The physical picture that emerges from our results indicates clearly the importance of the microscopic details of water flowing through hydrophobic nanochannels.
Notice that, we found that for large channel heights (H$>30\AA$), the well-known contribution of capillary pressure dominates in Eq.~(\ref{QQ}) where $f(z)\cong$ 1 and $g(z')\cong$1, which yields a linear dependence $Q\propto$ H. Moreover in order to provide an independent check on the chosen slip length in our study, we can roughly estimate  the slip length as follows. The Navier slip length is defined as $\lambda=\frac{\eta}{\xi}$ where $\xi$ is the water-solid interfacial friction coefficient. We calculated $\xi$ using the method introduced in Ref. [54], and for H$>$10\,\AA~found it almost independent of H, i.e. $\xi=3\times 10^4 Nsm^{-3}$. Using the obtained number for $\xi$ and those we already found for $\eta$, the slip length is in the range [500-700]\AA.~ All numbers in this range give a reasonable peak for Q around H=13\,\AA~ and give  different Q with only 1$\%$ difference. Therefore the  value $\lambda=600$\,\AA~as a mean value used in our analysis is reasonable.
}
 %However, the peak at H =11-15\AA~ arises from the rapidly rising disjoining pressure which is due to the crystal structure {of the %formed two water layers close to the  walls and their induced larger density $\rho$} which increases with decreasing  size of the capillary and %leads to an entropy term in the electrochemical potential and results in an extra pressure.

{Our modelling highlights the unique role of confinement on water flow for channel size H$\approx$13\AA~ and we found our results in very good agreement with the experiment of Ref. [20]. Below H=13\,\AA,~ the molecular regime dominates i.e. two layers of water are weakly bound to the walls of the channel and the available cross sectional flow area that remains between these two layers is much smaller than the molecular size of water leading to negligible flow. Around 13\,\AA,~ as shown in Fig. 2(d), a third layer of water appears between the two water layers which assemble into a two-dimensional structure for which neither the slip length nor the effective viscosity is well defined.  Water molecules in the middle layer are interacting weakly with the other layers and the walls, and their density is slightly smaller than the one of bulk water. However,  the total density of water inside the channel is larger than the bulk density. The water molecules in the middle layer diffuse freely. For larger H, although the two side layers of water are still present, the water molecules in the middle layer (having bulk density) randomly diffuse in the remaining space which results in a resistance against water flow and consequently a decrease in the flow rate.
  Note that the large viscosity and density we introduced in our analytical model  is for water inside the whole of the channel. If we subtract the contribution of the two water layers adjacent to the walls, both density and viscosity are only slightly smaller than the one of bulk for channels of size around 13\,\AA~and approach the bulk values for larger H. Notice that as shown in Fig. 3, both density and viscosity for the sub-continuum regime (H$\approx$13\,\AA)~ and the continuum regime (H$>$13\,\AA)~ are about the bulk values. This clearly indicates that for H$<$\,13\AA~ the effect of the two layers adjacent to the walls are important and they are effectively included by having a large density  and viscosity for H$<$13\,\AA.~Therefore, the peak at H =12-14\AA~ arises from the rapidly rising disjoining pressure which is due to the crystal structure of the formed two water layers close to the  walls and their induced larger density $\rho$.}

%The final decrease in $Q$ (for small channel heights (H$<11\AA$)) is due to the strong increase in viscosity which reduce the mass flow overtaking the entropic enhancement in capillary pressure.}

%In summary, our study uncovers an unexplored domain in nanofluidics with profound implications both for our fundamental understanding and for %technological applications of water flow through nanochannels.

We explained the observed  fast water flow through graphene nanochannels, which has been reported in a recent experiment\cite{Radha}, which finds its origin in the large density and the large viscosity of water inside nanochannels. Our MD simulations confirm the ordered structure and the large density and viscosity of confined water between graphene channels.
~~~~
~~~

{{{Acknowledgments.}}}
We acknowledge fruitful discussion with Andre K.
Geim and I. V. Grigorieva. This work was supported by the
Flemish Science Foundation (FWO-Vl) and the Methusalem
program. B.R. acknowledges the Royal Society Fellowship,
the L$’$Oreal fellowship for women in science, and EPSRC
Grant No. EP/R013063/1.


\begin{thebibliography}{15}
%\bibitem{ras} J. C. Rasaiah, S. Garde, and G. Hummer, Annu. Rev. Phys. Chem. {\bf59}, 713 (2008).

\bibitem{hendi}S. K. Kannam, B. D. Todd, J. S. Hansen, and P. J. Daivis, J. Chem. Phys. {\bf138}, 094701 (2013).
\bibitem{PNAS17} K. Wu, Z. Chen, J. Li, X. Li, J. Xu, and X. Dong, PNAS {\bf114}, 3358 (2017).
\bibitem{rev}D. Mattiaand Y. Gogotsi. Microfluid. Nanofluid. {\bf5}, 289 (2008).
\bibitem{falk}K. Falk, F. Sedlmeier, K. Joly, R. R. Netz, L.  Bocquet. Nano Lett. {\bf10}, 4067 (2010).
\bibitem{maj} M. Majumder, N. Chopra, R. Andrews, and B. J. Hinds, Nature {\bf438}, 44 (2005).
\bibitem{Holtsci}J. K. Holt, H. G. Park, Y. Wang, M. Stadermann, A. B. Artyukhin, C. P. Grigoropoulos, A. Noy, and O. Bakajin, Science {\bf19}, 5776 (2006).
\bibitem{Holt}  J. K. Holt, Adv. Mater. {\bf21}, 3542 (2009).
\bibitem{Qin} X. Qin, Q. Yuan Q, Y. Zhao, S. Xie, and Z. Liu, Nano Lett. {\bf11},  2173 (2011).
\bibitem{hummer} J. K\"{o}finger and G. Hummer, and C. Dellago, PNAS {\bf105}, 13218 (2008).
\bibitem{Thomas} J. A. Thomas and A. J. H. McGaughery, Nano Lett. {\bf8}, 2788 (2008).
\bibitem{Rasaiah} J. C. Rasaiah,  S. Garde, and D. Hummer,  Ann. Rev. Phys. Chem. {\bf59}, 713 (2008).
\bibitem{Lac} L. Lacerda, S.  Raffa, M. Prato, A. Bianco, and K.  Kostarelos.  Nano Today {\bf2}, 38 (2007).
\bibitem{Noy} A. Noy.  Nano Today {\bf2}, 22 (2007).
\bibitem{ber} A. Berezhkovskii and G. Hummer. Phys. Rev. Lett. {\bf89}, 064503 (2002).
\bibitem{jos} S. Joseph and N. R.  Aluru. Phys. Rev. Lett. {\bf101}, 064502 (2008).
\bibitem{thom} J. A. Thomas and A. J. H. McGaughey. Phys. Rev. Lett. {\bf102}, 184502 (2009).
\bibitem{alur}S. Joseph and N. R. Aluru, Phys. Rev. Lett. {\bf101}, 064502 (2008).
\bibitem{sokhan} V. P. Sokhan and N. Quirke, Phys. Rev. E {\bf78}, 015301R (2008).
\bibitem{quir} M. Whitby and N. Quirke, Nature Nanotechnol. {\bf2}, 87 (2007).

\bibitem{Radha}B. Radha, A. Esfandiar, F. C. Wang, A. P. Rooney, K. Gopinadhan, A. Keerthi, A. Mishchenko, A. Janardanan, P. Blake, L. Fumagalli, M. Lozada-Hidalgo, S. Garaj, S. J. Haigh, I. V. Grigorieva, H. A. Wu, and A. K. Geim, Nature (London) {\bf538}, 222 (2016).\bibitem{Israelachvili} J. N. Israelachvili, \emph{Intermolecular and Surface Forces,} 3rd ed., (Elsevier Academic, New York, 2011).
%\bibitem{Honschoten} J. W. van Honschoten, N. Brunets, and N. R. Tas, Chem. Soc. Rev. {\bf39}, 1096 (2010).
\bibitem{IEE}C. Mathew Mate, IEEE Transactions on Magnetics {\bf47}, 124 (2011).
\bibitem{PRE16} S. Gravelle, C.  Ybert, L. Bocquet, and  L. Joly, Phys. Rev. E {\bf93}, 033123 (2016).
%\bibitem{Kannam} S. K. Kannam, B. D. Todd, J. S. Hansen, and  P. J. Daivis, J. Chem. Phys. {\bf135} 144701 (2011).
\bibitem{Ahn} C. H. Ahn, Y. Baek, C. Lee, S. O. Kim, S. Kim, S. Lee, S.-H. Kim, S. S. Bae, J. Park, and J. J. Yoon, Ind. Eng. Chem. {\bf18}, 1551 (2012).

\bibitem{kh} M. Khademi and M. Sahimi. J. Chem. Phys. {\bf135}, 204509 (2011).
\bibitem{khh} M. Khademi, R. K. Kalia, and M.  Sahimi. Phys. Rev. E {\bf92}, 030301 (2015).
\bibitem{khhh}M. Khademi and  M. Sahimi. J. Chem. Phys. {\bf145}, 024502 (2016).
\bibitem{Ebrahimi} F. Ebrahimi, F. Ramazani, and M. Sahimi. Scientific Reports {\bf8},  7752 (2018).

\bibitem{Algara} G. Algara-Siller, O. Lehtinen, F. C. Wang, R. R. Nair, U. Kaiser, H. A. Wu, A. K. Geim, and  I. V. Grigorieva, Nature (London) {\bf519}, 443 (2015).
\bibitem{Qiu} H. Qiu, X. C. Zeng, and W. Guo, ACS Nano {\bf9}, 9877 (2015).
\bibitem{ACS16} M. Neek-Amal, F. M. Peeters, I. V. Grigorieva, and A. K. Geim, ACS Nano {\bf9}, 3685 (2016).
%\bibitem{note3}


    \bibitem{refnote3} Jesper S. Hansen, Jeppe C. Dyre, Peter Daivis, Billy D. Todd, and Henrik Bruus, Langmuir {\bf31} (49), 13275 (2015).
\bibitem{Sobrino} M. Sobrino Fernandez, M. Neek-Amal, and  F. M. Peeters, Phys. Rev. B {\bf92}, 245428 (2015).
\bibitem{Sobrino2} M. Sobrino Fernández, F.M. Peeters, and  M. Neek-Amal, Phys. Rev. B {\bf94}, 04546 (2016).
%\bibitem{natureRahul} K. S. Vasu, E. Prestat, J. Abraham, J. Dix, R. J. Kashtiban, J. Beheshtian, J. Sloan, P. Carbone, M. Neek-Amal, S. J. Haigh, A. K. Geim, and  R. R. Nair, Nature Communications {\bf7}, 12168 (2016).
%

%\bibitem{StanlyNatPhys} S. Han, M. Y. Choi, P. Kumar, and H. Eugene Stanley, Nat. Phys. {\bf6}, 685 (2010).
\bibitem{Zangi} R. Zangi and A.  E. Mark, J. Chem. Phys. {\bf120}, 7123 (2004).
\bibitem{Walt} Shun Chen, Adam Paul Draude, Xuechuan Nie, Haiping Fang, Niels R Walet, Shiwu Gao, and Jichen Li, to appear in J. Phys. Communications (2018).
\bibitem{Corsetti1} F. Corsetti, J. Zubeltzu, and E. Artacho, Phys. Rev. Lett. {\bf116}, 085901 (2016).
%\bibitem{Corsetti2} F. Corsetti, P. Matthews, and E. Artacho, Sci. Rep. {\bf6}, 18651 (2016).

\bibitem{hamid} H. Mosadeghi, S. Alavi, M. H. Kowsari, and Bijan Najafi, J. Chem. Phys. {\bf137}, 184703 (2012).
\bibitem{Pradeep}  P. Kumar, S. V. Buldyrev, F. W. Starr, N. Giovambattista, and H. Eugene Stanley, Phys. Rev. E. {\bf72}, 051503 (2005).
\bibitem{PRL2009}John A. Thomas and Alan J. H. McGaughey, Phys. Rev. Lett {\bf102}, 184502, (2009).
\bibitem{size1}H. Ye, H. Zhang, Z. Zhang, and Y. Zheng. Nanoscale Res. Lett. {\bf6}, 87 (2011).
 \bibitem{size2}J. A. Thomas, International Journal of Thermal Sciences {\bf49}, 281 (2010).
  \bibitem{size3}F. Calabrò, K. P. Lee, and D. Mattia. Applied Mathematics Letters {\bf26}, 991 (2013).

%\bibitem{Nature2001}U. Raviv, P. Laurat, and J. Klein, Nature (London), {\bf413}, 51 (2001).
%\bibitem{PRL1}S. H. Khan, G. Matei, S. Patil, and P. M. Hoffmann, Phys. Rev. Lett. {\bf105}, 106101 (2010).
%\bibitem{PRL2} T. D. Li and E. Riedo, Phys. Rev. Lett. {\bf100}, 106102 (2008).
%\bibitem{JACS} A. Dhinojwala and S. Granick, J. Am. Chem. Soc. {\bf119}, 241 (1997).
%\bibitem{MITreview} Eric Lauga, Michael P. Brenner and Howard A. Stone,   \emph{Handbook of Experimental Fluid Dynamics}, C. Tropea, A. Yarin, J. F. Foss (Eds.) Springer, New-York (2007) Chapter 19, pages 1219 - 1240.




%\bibitem {hamid} H. Mosaddeghi, S. Alavi,  M. H. Kowsari,  and  B. Najafi, J. Chem. Phys., {\bf137}, 184703 (2012).
\bibitem{slabgeom} H. Itoh and H. Sakuma, J. Chem.  Phys. {\bf142}, 184703 (2015).
\bibitem{surfacetension} C. T Nguyen and B. Kim, International J. of Precision Engineering and Manufacturing {\bf17}, 503 (2016).

\bibitem{Taherian} F. Taherian, V. Marcon, and N. F. A. van der Vegt, Langmuir {\bf29}, 1457 (2013).
\bibitem{Mucksch} C. M\"{u}cksch, C. R\"{o}sch, C. M\"{u}ller−Renno, C. Ziegler, and H. M. Urbassek, J. Phys. Chem. C {\bf119}, 12496 (2015).
\bibitem{zisman}W. A. Zisman, Advances in Chemistry, {\bf43}, 1-51 (1964).
\bibitem{nanolett2001}T. Werder, J. H. Walther, R. L. Jaffe, T. Halicioglu, F. Noca, and P. Koumoutsakos, Nano Letters {\bf1} 697, (2001).
\bibitem{DP2}  C. M. Mate, IEEE Trans. Magn. {\bf47}, 124, (2011); B. V. Derjaguin and N. V. Churaev, Journal of Colloid and Interface Science {\bf49}, 249 (1974).

%\bibitem{scaling} {This parameter, when using experimental numbers $w=$1300\AA~ and $L=10^4$\AA~for one channel, is about 1.2172$\times10^{-4}$ s\AA$^{-2}$ which is very close to our obtained number from a fitting on the experimental data, i.e. 1.2174$%%%\times10^{-4}$ s\AA$^{-2}$.}

\bibitem{tip3p} W. L. Jorgensen,  J. Chandrasekhar,  J. D. Madura, R. W. Impey,  and  M. L. Klein, J. Chem. Phys. {\bf79}, 926 (1983).
\bibitem{arxiv2017} J. Zubeltzu and E. Artacho, arXiv:1705.05270 (2017).
%\bibitem{note1} Here we assume that the same power for exponential terms, i.e $e^{-z}$, for both viscosity and density. We do this assumption to have a few numbers of free parameters in our model.

%\bibitem{note4} We use $A_0$  as another free parameter we fit Eq.~(\ref{QQ}) on the experimental data.~\cite{Radha}

%\bibitem{note} The graphene layers are kept rigid during simulations. At the start of the simulations, in order to fill the  nanochannels, an additional graphene layer (perpendicular to the channel) pushes water inside the channel. {Periodic boundary conditions are applied in %yz-plane inside the reservoirs and along y-direction inside the channel.} The  temperature was kept constant at room temperature using Nos'e-Hoover thermostat.  {The maximum estimated lateral pressure for filling the channel is about 0.9GPa (for H=6.5\,\AA)~which %decreased with H}. In fact, we wait for thermodynamic equilibrium when the channels are filled and the density of the reservoirs are the bulk density (which provides us the  density of water inside the channels $\rho_0$). In order to provide an independent check for these %observed effects, we performed several additional MD simulations using different force fields such as ReaxFF\cite{reaxff} and TIP4P\cite{tip4p}. We found that independent of the employed force field all of the performed MD simulations resulted in large water densities for %H$\preceq12$\AA.~
%\bibitem{note4} Notice that the ordered structure and large density of water is well known for the cubic ice (ice VII) structure\cite{vii} which is formed under  3\,GPa pressure having 1.65\,gcm$^{-3}$density.
%\bibitem{reaxff} A. C. T. van Duin, S. Dasgupta, F. Lorant, and W. A. Goddard, J. Phys. Chem. A. {\bf105}, 9396 (2001); A. C. T. van Duin and J. S. S. Damste, Org. Geochem. {\bf34}, 515 (2003).

   % \bibitem{tip4p}W. L. Jorgensen, J. Chandrasekhar, and J. D. Madura, J. Chem. Phys. {\bf79}, 926 (1983).
   % \bibitem{vii} D. Eisenberg and W. Kauzmann, \emph{The structure and properties of water} (Oxford University Press, London, 1969); http://www1.lsbu.ac.uk/water.

%Notice that one may use the vdW radius of O and C atoms for defining the the volume distance, but, this may not bring the whole physics because the usually used vdW  radius are defined for water at conditions (under a pressure of 1\,at). Nevertheless, even using vdW radius  give large density at GPa order pressure. For instance, for the pressures around 0.9GPa, and by considering the vdW radius issue, Kumar et al.~\cite{Pradeep} found a density of about 1.25gcm$^{-3}$ for H=11.\AA.
\bibitem{Hamaker} J.-L Li, J. Chun, N. S. Wingreen, R. Car, I. A. Aksay, and D. A. Saville, Phys. Rev. B. {\bf71}, 235412 (2004).
%~RADHA CHECK THIS PLEASE}
\bibitem{eta}Bladimir Ramos-Alvarado, Satish Kumar, and G. P. Peterson, Phys. Rev. E {\bf93} 023101 (2016).
%\bibitem{note1} The Navir slip length is defined as $ \lambda=\frac{\eta}{\xi}$\cite{ACS16} where $\xi$ is the water-solid interfacial friction coefficient\cite{}. Using dependence of viscosity on H ($g(z')$) and approximating friction coefficient by\cite{sokhan} $\xi\cong \frac{H \rho(H)}{\tau}$, we were able to calculate the the slip length where found to be in the range $\lambda\in$ [500-1000]\,\AA.~

\end{thebibliography}
\end{document}